\documentclass[12pt,a4paper]{article}

\usepackage{authblk}
\usepackage{amsmath,amssymb}
\usepackage{graphicx}
\usepackage{amsfonts}
\usepackage{epsfig}

\usepackage{tikz}
\usepackage{pgflibrarysnakes}
\usetikzlibrary[snakes]

\usetikzlibrary{arrows,shapes,positioning}
\usetikzlibrary{decorations.markings}
\tikzstyle arrowstyle=[scale=1]
\tikzstyle directed=[postaction={decorate,decoration={markings,
		mark=at position .65 with {\arrow[arrowstyle]{stealth}}}}]
\tikzstyle reverse directed=[postaction={decorate,decoration={markings,
		mark=at position .65 with {\arrowreversed[arrowstyle]{stealth};}}}]

\renewcommand{\baselinestretch}{1.5}

\newcommand{\im}{\mathrm i}

\newcommand{\tr}{\operatorname{Tr}}

\newcommand{\eq}{\begin{equation}}
\newcommand{\en}{\end{equation}}
\newcommand{\bear}{\begin{eqnarray}}
\newcommand{\ear}{\end{eqnarray}}

\title{\mbox{}On the partition function of the $Sp(4)$ integrable vertex model} 

\author{G.A.P. Ribeiro\footnote{E-mail: pavan@df.ufscar.br}}
\affil{\small Departamento de F\'isica, Universidade Federal de S\~ao Carlos, PO Box 676, 13565-905, S\~ao Carlos-SP, Brazil\vspace{8pt}}
\author{A. Kl\"umper}
\affil{\small Theoretische Physik, Bergische Universit\"at Wuppertal\\ 42097
Wuppertal, Germany\vspace{8pt}}
\author{P.A. Pearce}
\affil{\small School of Mathematics and Statistics, 
	University of Melbourne\\ 
	Parkville, Victoria 3010, Australia}
\affil{\small School of Mathematics and Physics, University of Queensland, St Lucia,
Brisbane, Queensland 4072, Australia}

\date{}

\begin{document}
\renewcommand{\baselinestretch}{1.2}
\maketitle

\vspace{-.5in}\noindent
\begin{abstract}
In this paper we investigate certain fusion relations associated to an integrable
vertex model on the square lattice which is invariant under $Sp(4)$ symmetry. We establish a set of functional 
relations which include a transfer matrix inversion identity. The solution
of these relations in the thermodynamic limit allows us to compute the partition function per site 
of the fundamental $Sp(4)$ representation of the vertex model. As a byproduct we also obtain
the partition function per site of a vertex model mixing the four and five dimensional representations of the $Sp(4)$ symmetry.
\end{abstract}

\centerline{Keywords: Integrability, vertex models, inversion relations, fusion }
\renewcommand{\baselinestretch}{1.5}

\thispagestyle{empty}

\newpage

\pagestyle{plain}

\pagenumbering{arabic}

\section{Introduction}

The study of statistical models on the square lattice via transfer matrix techniques has a long history \cite{BAXTER,BOOK-KBI}. Over the years, the use of approaches based on functional equations has been shown to be fruitful in dealing with models of intricate algebraic or analytical structure. More specifically, in the context of vertex models and IRF models,  one has approaches based on inversion relations	\cite{STROGANOV,BAXTER-inversion,SHANKAR}, fusion functional equations~\cite{BAXTER-PEARCE,KIRILLOV,BAZHANOV} and transfer matrix inversion identities~\cite{PEARCE1987}.

In this work, we are interested in the determination of the partition function of the $Sp(4)$ integrable vertex model via functional methods, which together with Bethe ansatz techniques and $T$-$Q$ relations are the state of the art of the computation of the partition functions.At the classical level, the $Sp(4)$ vertex model is interesting (along with the general $Sp(2n)$ and $O(n)$ vertex models) since, along with the identity and permutation operators, it involves  the Temperley-Lieb operators and so generalizes the $SU(n)$ vertex models. On the other hand, the $Sp(4)$ quantum spin chain can be seen as coupled spin-$1/2$ chains in the context of spin-orbital models \cite{SPIN-ORBITAL}. Besides that, the interest in the $Sp(4)$ model is enhanced due to the subtle analytical structure of the leading eigenvalue, which implies that one cannot 
straightforwardly solve the usual inversion relation. In order to proceed, we need to complement the usual inversion relation with suitable additional relations obtained from the rules of fusion \cite{KAROWSKI,KULISH1981,KULISH1982,CAO}. We derive a set of relations that we refer to as transfer matrix fusion identities which hold for arbitrary values of the spectral parameter. This is in contrast with the discrete set of relations used in \cite{CAO}.

In the thermodynamic limit, the above mentioned transfer matrix fusion identities are an exact truncation of the fusion hierarchy, which therefore allows for the complete determination of the partition function per site of the vertex model on the square lattice. The obtained solution can be seen to show explicitly a kind of CDD factor due to the loss of analyticity along an infinitely long line in the complex plane. This seems to be the first instance of rational model presenting such features.

Moreover, the determination of the partition function of the transfer matrix in the thermodynamic limit is the first step towards the determination of the two-sites correlation function of the $Sp(4)$ quantum spin chain, along the same lines as \cite{RIBEIRO}. Similarly to \cite{RIBEIRO}, a set of two functional equations for the physical correlations can be obtained from the quantum Knizhnik-Zamolodchikov equations \cite{BOOS05,BOOS2} for the  $Sp(4)$ spin chain. Nevertheless, one of these equations can be cast into the form of the inversion relbation discussed in this work. Therefore, due to the loss of analyticity in the putative physical strip, it is necessary to enlarge the system of equations to fully determine this first function. The general framework to work out, from the quantum Knizhnik-Zamolodchikov equations, namely the full set of generalized fusion equations to fully determine the correlation functions is still an open problem.

This paper is organized as follows. In section \ref{INTEGRA}, we outline the integrable structure of the model. In section~\ref{fusion}, we introduce the fusion properties and obtain the transfer matrix fusion identities for the $Sp(4)$ model. In section~\ref{limit}, we compute the partition function per site in the thermodynamic limit. Our conclusions are given in section~\ref{CONCLUSION}. Additional details are given in two appendices.

\section{The vertex model}\label{INTEGRA}

The fundamental $Sp(4)$ integrable vertex model is described by the $R$-matrix \cite{RESHETIKHIN,KUNIBA,KULISH,MARTINS1997}
\eq
R_{12}^{(4,4)}(\lambda)= \lambda (\lambda+3) I_{12} + (\lambda+3) P_{12} +\lambda E_{12}.
\label{Rmatrix}
\en
which acts in the indicated spaces of the tensor product $W\otimes W$, 
where $W$ is the fundamental representation of $Sp(4)$, which is  of dimension $4$ as indicated in the superscript. Here $I_{i,i+1}$, $P_{i,i+1}$ and $E_{i,i+1}$ are the identity, permutation and Temperley-Lieb operators acting on the sites $i$ and $i+1$.  Their matrix elements are given as $(I_{i,i+1})_{ac}^{bd}=\delta_{a,b}\delta_{c,d}$, $(P_{i,i+1})_{ac}^{bd}=\delta_{a,d}\delta_{b,c}$ and $(E_{i,i+1})_{ac}^{bd}=\epsilon_{a} \epsilon_c \delta_{a,5-c}\delta_{b,5-d}$ for $1\leq a,b,c,d \leq 4$ where $\epsilon_1=\epsilon_2=1$ and $\epsilon_3=\epsilon_4=-1$. 

The $R$-matrix has the important properties of regularity, unitarity and crossing as given below,
\bear
R_{12}^{(4,4)}(0)&=& 3 P_{12}, \\
R_{12}^{(4,4)} (\lambda) R_{21}^{(4,4)} (-\lambda) &=& (1-\lambda^2) (3^2-\lambda^2)I_{12}, \\
R_{12}^{(4,4)} (\lambda) &=& (V\otimes I) (R^{(4,4)}_{12}(-\lambda-\rho) )^{t_2} (V^{-1}\otimes I),
\ear
where $t_2$ is transposition in the second space, the crossing parameter is $\rho= 3$ and the crossing matrix $V$ is given by $V=\mbox{anti-diagonal}(1,1,-1,-1)$, where the matrix entries are listed from the top-right to the bottom-left corners. The $R$-matrix satisfies the Yang-Baxter equation
\eq
R^{(4,4)}_{12}(\lambda-\mu) R^{(4,4)}_{13} (\lambda) R^{(4,4)}_{23} (\mu) =R^{(4,4)}_{23}(\mu) R^{(4,4)}_{13}(\lambda)  R^{(4,4)}_{12} (\lambda-\mu).
\label{yang-baxter}
\en

The partition function of a classical vertex model on an $M\times L$ lattice with periodic boundary conditions in both directions can be written as $Z=\tr{\left[\left(T^{(4)}(\lambda)\right)^M\right]}$, where $T^{(4)}(\lambda)$ is the row-to-row transfer matrix given by the trace over the $4$-dimensional auxiliary space $\cal A$  of the monodromy matrix  ${\cal T}_{\cal A}^{(4,4)}(\lambda)=R_{{\cal A}L}^{(4,4)}(\lambda)\dots R_{{\cal A} 1}^{(4,4)}(\lambda)$ such that,
\eq
T^{(4)}(\lambda)=\tr_{\cal A}{[ {\cal T}_{\cal A}^{(4,4)}(\lambda)]}.
\en
Thanks to the Yang-Baxter equation, the transfer matrix constitutes a family of commuting operators with \mbox{$[T^{(4)}(\lambda),T^{(4)}(\mu)]=0$}. Therefore, $T^{(4)}(\lambda)$ is a generating function of conserved charges. The first non-trivial conserved charge is obtained by logarithmic derivative of the transfer matrix, ${\cal H}^{(4)}=\frac{d}{d\lambda}\log{T^{(4)}(\lambda)}\Big|_{\lambda=0}$, which is the Hamiltonian  of the integrable $Sp(4)$ spin chain with periodic boundary condition \cite{RESHETIKHIN,KUNIBA,KULISH,MARTINS1997,MARTINS,BATCHELOR,MARTINS-SP2N},
\eq
{\cal H}^{(4)}=\sum_{i=1}^{L}\left(\frac{1}{3}I_{i,i+1} +P_{i,i+1} - \frac{1}{3} E_{i,i+1}\right),
\en
whose physical properties were studied via the solution of the Bethe ansatz equation in \cite{MARTINS-SP2N}.

The above Hamiltonian can be written in terms of two commuting sets of Pauli matrices \cite{BATCHELOR,MARTINS-SP2N} and it is related to spin-orbital models \cite{SPIN-ORBITAL},
\eq
{\cal H}^{(4)}=\frac{1}{6}\sum_{i=1}^{L}\left[ \vec{\sigma}_i \vec{\sigma}_{i+1} +2 (\vec{\tau}_i \vec{\tau}_{i+1} -\tfrac{1}{2} \tau_i^z \tau_{i+1}^z ) + \vec{\sigma}_i \vec{\sigma}_{i+1}(\vec{\tau}_i \vec{\tau}_{i+1} + \tau_i^z \tau_{i+1}^z) + 4 I_{i,i+1} \right],
\en
where $\vec{\sigma}_i$ is associated with the spin degrees of freedom and  $\vec{\tau}_i$ represents the orbital degree of freedom.

\section{Fusion relations}\label{fusion}

For the $Sp(4)$ case, the tensor product of two fundamental representations decomposes as $4\otimes 4 = 1 \oplus 5 \oplus 10$~\cite{GROUP}. 
This means that we can rewrite the fundamental
$R$-matrix $R^{(4,4)}_{12}(\lambda)$ in terms of the projectors onto such spaces, namely
\eq
R^{(4,4)}_{12}(\lambda)=(\lambda+1)(\lambda-3) \check{P}_{12}^{(1)} + (\lambda-1)(\lambda+3) \check{P}_{12}^{(5)}  + (\lambda+1)(\lambda+3) \check{P}_{12}^{(10)}, 
\en
where $\check{P}_{12}^{(\alpha)}$ are the projectors on the  $\alpha$-dimensional subspaces ($\alpha=1,5,10$), which are given in Appendix~A. This shows explicitly the singular $\lambda$ values for which the $R$-matrix degenerates into projection operators and implies $R^{(4,4)}_{12}(-3) =12 \check{P}_{12}^{(1)}$ and $R^{(4,4)}_{12}(-1) =-4 \check{P}_{12}^{(5)}$.

By the rules of fusion \cite{KAROWSKI,KULISH1981,KULISH1982}, one can exploit the point $\lambda=-1$ to obtain a new $R$-matrix with a $5$-dimensional auxiliary space (see \cite{CAO,ODBA} and the Appendix~A for more details on the fusion rules for $Sp(4)$) given as,
\eq
R_{12}^{(5,4)}(\lambda)=(\lambda-\frac{5}{2}) \check{P}_{12}^{(4)} + (\lambda+\frac{5}{2})\check{P}_{12}^{(16)},
\en
where the projectors $\check{P}_{12}^{(4)}$ and $\check{P}_{12}^{(16)}$ (also given in Appendix~A) are due to the decomposition $5\otimes 4 = 4 \oplus 16 $. From the singular values, one can also simply read that $R_{12}^{(5,4)}(\frac{5}{2})=5 \check{P}_{12}^{(16)}$ and $R_{12}^{(5,4)}(-\frac{5}{2})=-5 \check{P}_{12}^{(4)}$. The latter value is of particular importance, since it has the potential of closing the fusion hierarchy by projecting back to the fundamental representation. Besides, this fused $R$-matrix also satisfies the unitarity condition $R_{12}^{(5,4)} (\lambda) R_{21}^{(4,5)} (-\lambda) = ((\frac{5}{2})^2-\lambda^2) I_{12}$ and the following Yang-Baxter equation,
\eq
R_{12}^{(5,4)}(\lambda-\mu) R_{13}^{(5,4)} (\lambda) R_{23}^{(4,4)} (\mu) =R_{23}^{(4,4)}(\mu) R_{13}^{(5,4)}(\lambda)  R_{12}^{(5,4)} (\lambda-\mu).
\label{yang-baxter2}
\en

Having this fused $R$-matrix, we can define another transfer matrix whose auxiliary space is $5$-dimensional,
\eq
T^{(5)}(\lambda)=\tr_{\cal A}{\left[R_{{\cal A}L}^{(5,4)}(\lambda) R_{{\cal A}L-1}^{(5,4)}(\lambda)\cdots R_{{\cal A}1}^{(5,4)}(\lambda)\right]}.         
\en
The above transfer matrices $T^{(4)}(\lambda)$ and $T^{(5)}(\lambda)$ also commute mutually for different spectral parameters.

The fusion structure provides us with the functional relations among these transfer matrices, which we refer to as transfer matrix fusion identities. For $Sp(4)$, these take the form
\begin{align}
T^{(4)}(\lambda)T^{(4)}(\lambda-3)&= [(\lambda^2-1) (\lambda^2-3^2)]^L I + \varphi(\lambda) V_1(\lambda), \label{TMFI1} \\
T^{(4)}(\lambda)T^{(4)}(\lambda-1)&= [(\lambda^2-1) (\lambda+3)]^L T^{(5)}(\lambda-\frac{1}{2}) + \varphi(\lambda) V_2(\lambda), \label{TMFI2}\\
T^{(4)}(\lambda)T^{(5)}(\lambda-\frac{5}{2})&= (\lambda+3)^L T^{(4)}(\lambda-2) + \varphi(\lambda) V_3(\lambda), \label{TMFI3}
\end{align} 
where (\ref{TMFI1}) is the transfer matrix inversion identity \cite{PEARCE1987}.  By exploiting the singular values $\lambda=-1$ and $\lambda=-\frac{5}{2}$ of $R_{12}^{(4,4)}(\lambda)$ and $R_{12}^{(5,4)}(\lambda)$ respectively, we can derive the two additional identities between the fusion transfer matrices (\ref{TMFI2}-\ref{TMFI3}). Here $\varphi(\lambda)=\lambda^L$ and $V_{i}(\lambda)$ are analytic auxiliary matrices that commute with the transfer matrices. Moreover,  $\varphi(0)=0$ so that (\ref{TMFI1}) is trivially satisfied when $\lambda=0$ due to the pure action of the projectors and (\ref{TMFI2}--\ref{TMFI3}) become special relations between the $4$ and $5$ dimensional transfer matrices, since $T^{(4)}(0)$ is the shift operator. We note that these identities are derived for periodic boundaries. Nevertheless by changing the boundary conditions, we would only introduce $O(1)$ terms into these equations that do not change the calculation of the bulk free energies. This is in agreement with the expectation that the bulk free energies are independent of the choice of boundary conditions that do not break any transfer matrix conservation laws.

Since the common eigenvectors of the transfer matrices and auxiliary matrices are independent of $\lambda$, the functional relations (\ref{TMFI1}-\ref{TMFI3}) are also satisfied by the transfer matrix eigenvalues $\Lambda^{(\alpha)}(\lambda)$ and the auxiliary matrix eigenvalues ${\cal V}_i(\lambda)$. In addition to that, for large $L$, the terms $\varphi(\lambda) {\cal V}_i(\lambda)$ are exponentially small when compared with the other terms, since it is enough to consider $|\lambda|<\epsilon$ for an arbitrarily small and positive $\epsilon$, which implies that those terms decay as $|\lambda|^L \sim \epsilon^L$ as is verified numerically and also by the consistency with the numerical estimated bulk free energy and the ground state energy of the associated spin chain. Therefore, the fusion identities reduce to, 
\begin{align}
\Lambda^{(4)}(\lambda)\Lambda^{(4)}(\lambda-3)&= [(\lambda^2-1) (\lambda^2-3^2)]^L\left( 1  + O(e^{-L})\right), \label{TMFR1} \\
\Lambda^{(4)}(\lambda)\Lambda^{(4)}(\lambda-1)&= [(\lambda^2-1) (\lambda+3)]^L \Lambda^{(5)}(\lambda-\frac{1}{2})\left( 1  + O(e^{-L})\right), \label{TMFR2}\\
\Lambda^{(4)}(\lambda)\Lambda^{(5)}(\lambda-\frac{5}{2})&= (\lambda+3)^L \Lambda^{(4)}(\lambda-2)\left( 1  + O(e^{-L})\right), \label{TMFR3}
\end{align}
which are valid for $|\lambda|<\epsilon<\tfrac{1}{2}$ for some fixed $\epsilon$. Note that (\ref{TMFR1}) is the inversion relation \cite{STROGANOV,BAXTER-inversion,SHANKAR}
for the $Sp(4)$ vertex model. Therefore, the above fusion relations are an exact truncation of the fusion hierarchy in the limit $L\rightarrow\infty$.

The above relations hold for all the transfer matrix eigenvalues for large $L$. In the limit $L\to\infty$, the eigenvalues are determined by their different analyticity and asymptotic behaviours within the analyticity strip. 
The analyticity strip is determined by the pattern of zeros of the eigenvalues and coincides with the largest strip which may contain a finite number of (removable) zeros but is free of accumulating zeros as $L\to\infty$. In this case, referring to the numerical data (see Figure 1 below), the analyticity strip is either $-\tfrac{7}{2} < \mbox{Re}(\lambda) < \tfrac{1}{2}$ or a somewhat narrower strip. 
To calculate free energies, it suffices to consider the region $-3-\epsilon < \mbox{Re}(\lambda) < \epsilon$, where $\epsilon$ is fixed within the interval $0<\epsilon<\tfrac{1}{2}$. Once a bulk free energy is obtained in an open domain inside the analyticity strip, it can be analytically continued to the full analyticity strip.

In order to determine the partition function of the fundamental model, we are interested in the largest eigenvalue $\Lambda_0^{(4)}(\lambda)$. In most cases, the leading eigenvalue 
is an analytic function in the respective physical strip with at most a finite number
of zeros which can be removed by the multiplication of a finite number of polynomial factors
usually called CDD factors in the context of (1+1)-dimensional field theories \cite{CDD}.
This allows for the determination of the leading eigenvalue by using the single inversion relation (\ref{TMFR1}) as in \cite{STROGANOV,BAXTER-inversion,SHANKAR} and also for the $O(n)$ vertex models \cite{KLUMPER1990}.

\begin{figure}[t]
	\begin{center}
		\begin{minipage}{0.5\linewidth}
		\includegraphics[width=0.65\linewidth, angle=-90]{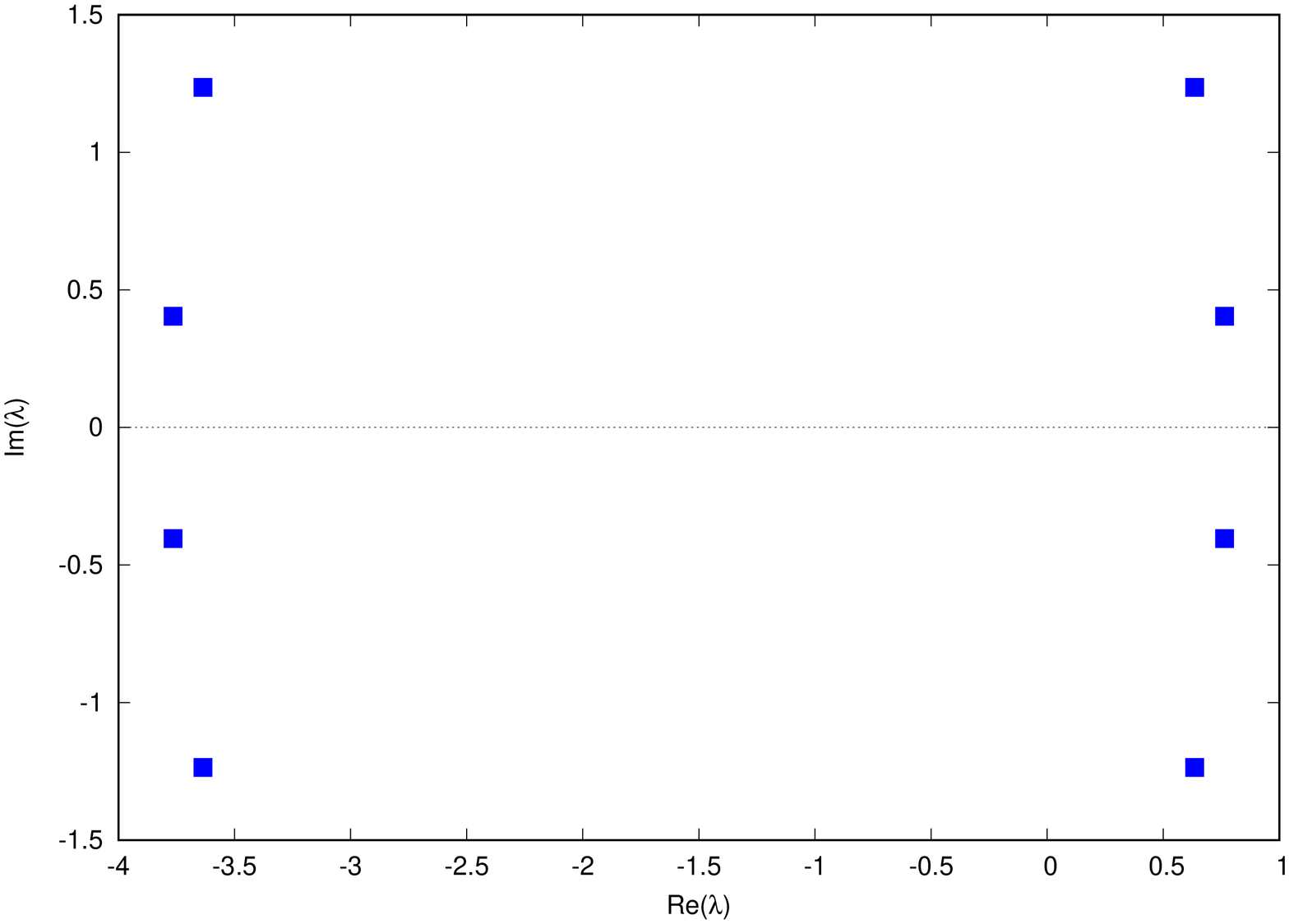}
		\includegraphics[width=0.65\linewidth, angle=-90]{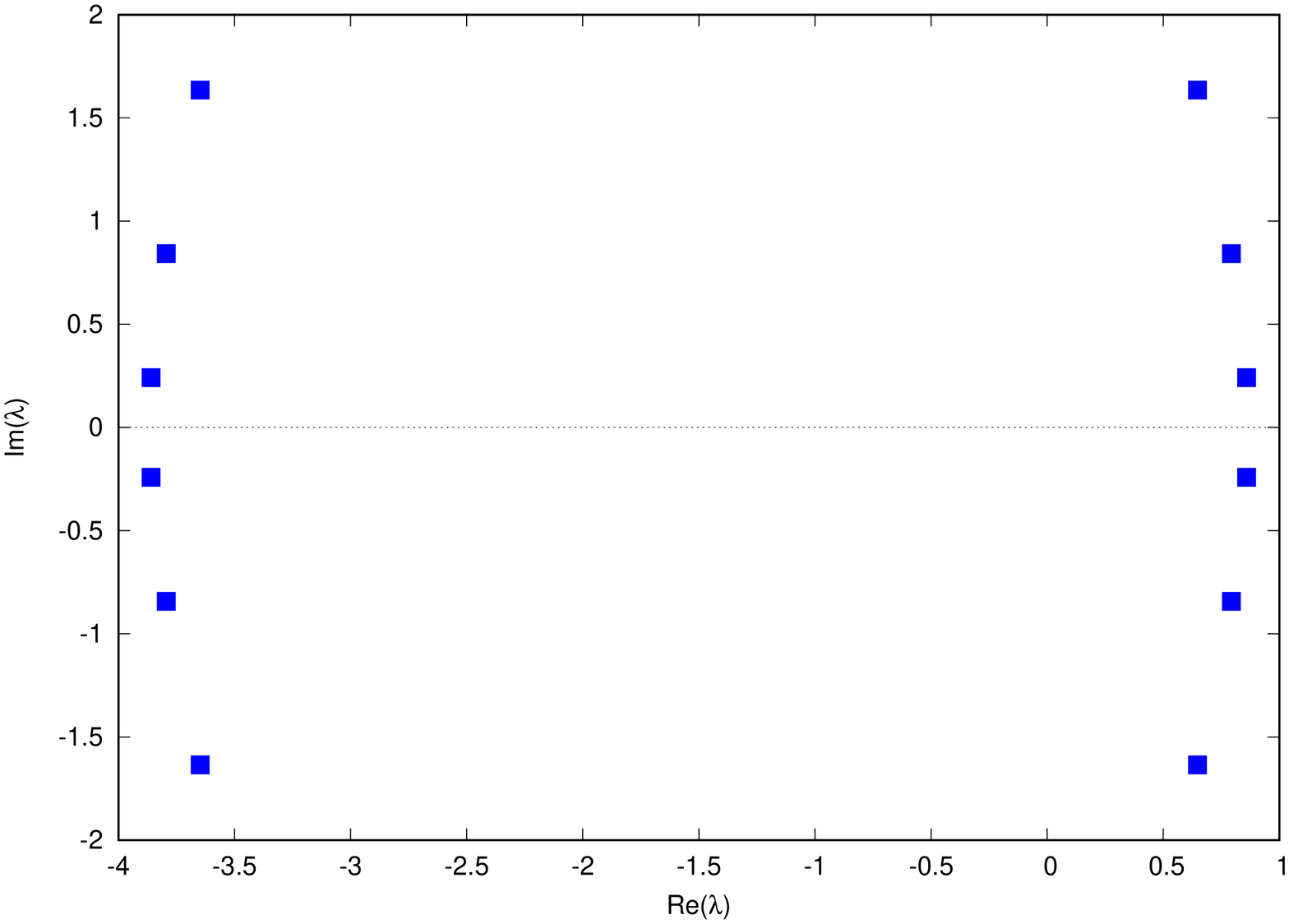}
		\end{minipage}%
        \begin{minipage}{0.5\linewidth}
		\includegraphics[width=0.65\linewidth, angle=-90]{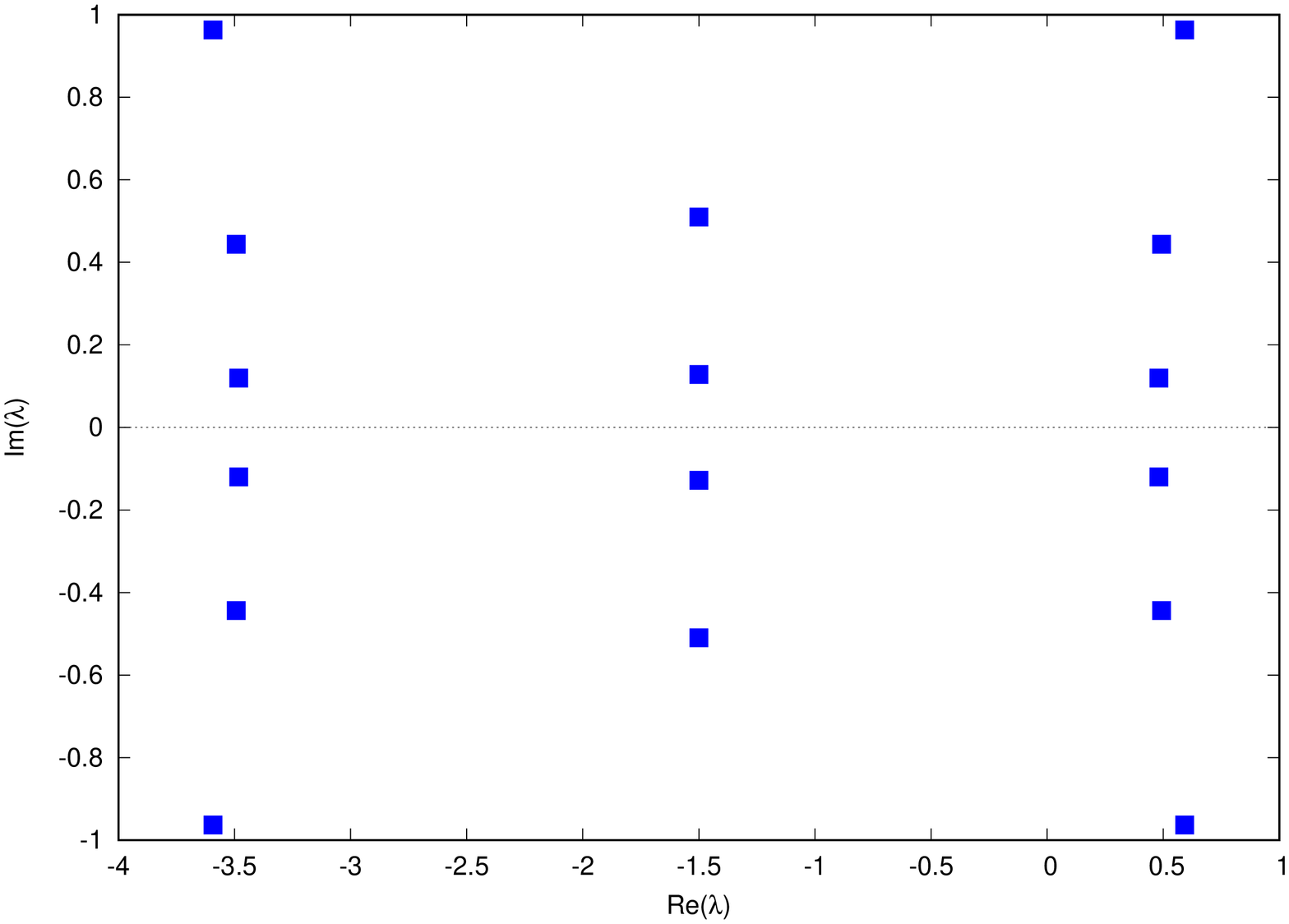}
		\includegraphics[width=0.65\linewidth, angle=-90]{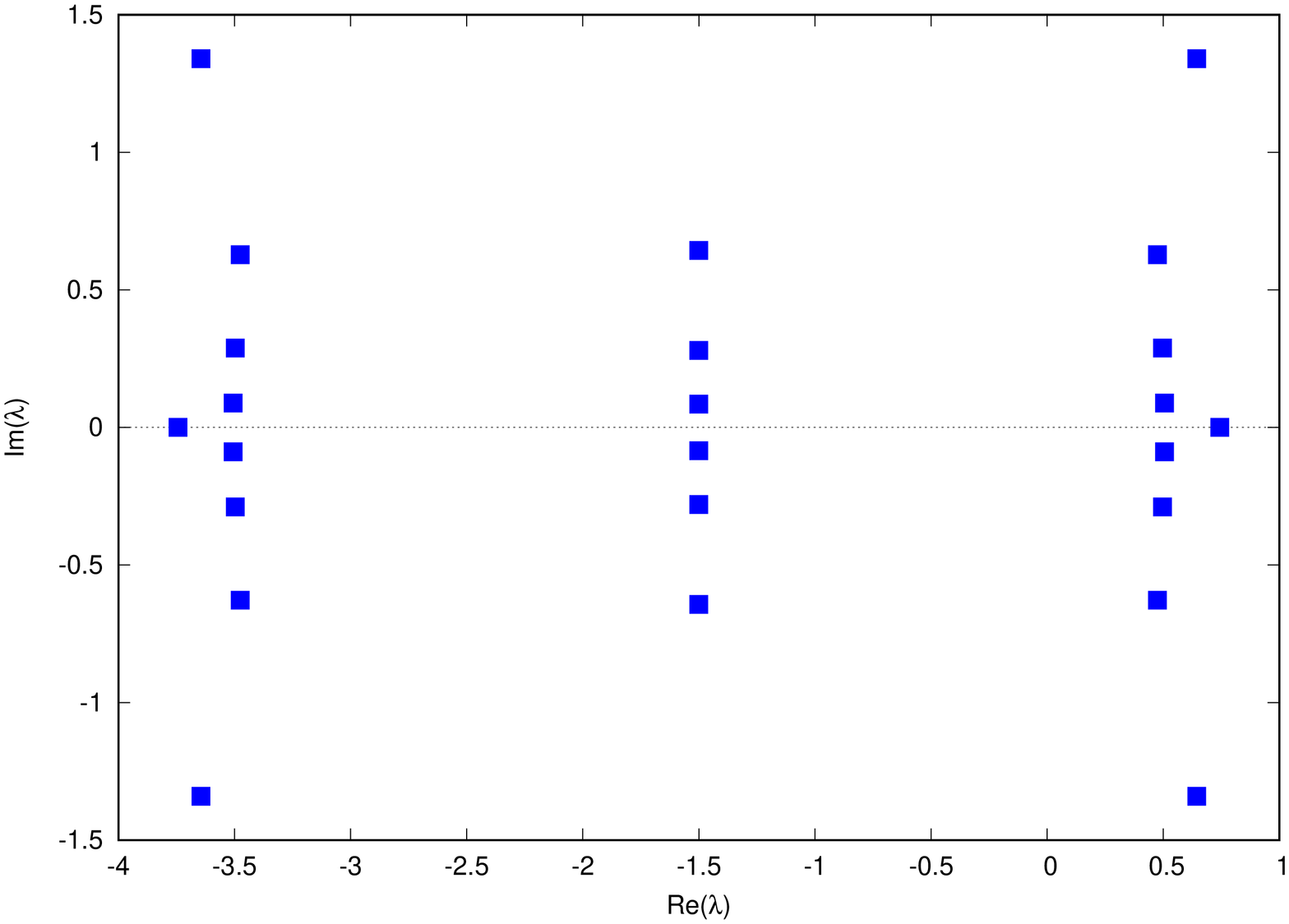}
		\end{minipage}
		\caption{Left panel: Zeros of the leading eigenvalue $\Lambda_0^{(5)}(\lambda)$ for $L=8,12$.
Right panel: Zeros of the leading eigenvalue $\Lambda_0^{(4)}(\lambda)$ in the complex plane for $L=8,12$ exhibiting $L/2$ zeros in the center of the analytical strip. }
		\label{fig1}
	\end{center}
\end{figure}

\def\Re{\mathop{\rm Re}}
We have studied the pattern of zeros of the largest eigenvalue of the transfer matrix for the $Sp(4)$ vertex model. Our results are exhibited in Figure~\ref{fig1} for lattice sizes
$L=8$ and $L=12$ respectively. This study reveals that, for the $Sp(4)$ vertex model, there are zeros precisely at the center of the strip along the line $\Re(\lambda)=-\frac{3}{2}$ and the number of those zeros and their density grows linearly with system size. This accumulation of zeros breaks the analyticity along the center line of the physical strip.
This means that the $Sp(4)$ model is a rather peculiar model in that it requires extra relations to connect both sides of the physical strip in order to determine the partition function in the thermodynamic limit. Since the largest eigenvalue of the transfer matrix with $5$-dimensional auxiliary space $\Lambda_0^{(5)}(\lambda)$ is free of zeros inside the wider strip $-\frac{7}{2}<\mbox{Re}(\lambda)<\frac{1}{2}$, the set of relations (\ref{TMFR1}--\ref{TMFR3}) are just enough to determine the leading eigenvalues $\Lambda^{(4)}_0(\lambda)$, $\Lambda_0^{(4)}(\lambda-3)$ and $\Lambda_0^{(5)}(\lambda)$. This is the subject of the next section.

It is worth noting that the exact truncation of the fusion hierarchy was exploited earlier in the different context of inhomogeneous transfer matrix, which is the core of the off-diagonal Bethe ansatz method \cite{CAO,ODBA}. In the framework of \cite{CAO,ODBA}, the fusion relations are truncated exactly at certain discrete values of the spectral parameter and the relations hold for arbitrary lattice sizes. We expect that the above exact truncation of the fusion hierarchy for the transfer matrix with periodic boundary condition in the thermodynamic limit can also be extended to the case of the double-row transfer matrix. This would require the use of the fused boundary $K$-matrices and the consideration of the role of reflecting monodromy matrix as described in \cite{CAO,ODBA}. Therefore, we expect that relations similar to (\ref{TMFI1}-\ref{TMFR3}) should exist for the double-row transfer matrix with the insertion of multiplicative factors due to the left and right boundaries. However the explicit derivation of these relations and calculation of the associated boundary free energies are beyond the scope of this work.

\section{Largest eigenvalue of the transfer matrix and partition function}\label{limit}

In this section we determine the largest eigenvalue $\Lambda_0^{(4)}(\lambda)$ of the transfer matrix in the extended region $-\frac{7}{2}<\mbox{Re}(\lambda)<\frac{1}{2}$. As a byproduct we also obtain the largest eigenvalues $\Lambda_0^{(5)}(\lambda)$.

In order to do this, it is convenient to define the partition function per site and its logarithmic derivative
\eq
\kappa^{(\alpha)}(\lambda)= \lim_{L\rightarrow\infty} \left(\Lambda_0^{(\alpha)}(\lambda)\right)^{1/L}\!\!, \quad\; 
\omega^{(\alpha)}(\lambda)=\frac{d}{d\lambda}\log{\kappa^{(\alpha)}}(\lambda),\quad\; \alpha=4,5.\label{kappaomega}
\en
Allowing for the break in analyticity, we further write
\eq
\kappa^{(4)}(\lambda)=\begin{cases}
\kappa^{(4)}_I(\lambda),&\lambda<-\tfrac{3}{2}\\
\kappa^{(4)}_{II}(\lambda),&\lambda>-\tfrac{3}{2}
\end{cases},\qquad
\omega^{(4)}(\lambda)=\begin{cases}
\omega^{(4)}_I(\lambda),&\lambda<-\tfrac{3}{2}\\
\omega^{(4)}_{II}(\lambda),&\lambda>-\tfrac{3}{2}
\end{cases}\label{break}
\en 
where the indices $I$ and $II$ specify the functions on the left and right of the cut line $\mbox{Re}(\lambda)=-\frac{3}{2}$ with $\omega_{i}^{(4)}(\lambda)=\frac{d}{d\lambda}\log{\kappa_i^{(4)}}(\lambda)$ for $i=I$ or $II$. 

Using these functions, the fusion relations (\ref{TMFR1}--\ref{TMFR3}) can be rewritten as
\begin{align}
\kappa_{II}^{(4)}(\lambda)\kappa_{I}^{(4)}(\lambda-3)&= (\lambda^2-1) (\lambda^2-3^2),  
\nonumber \\
\kappa_{II}^{(4)}(\lambda)\kappa_{II}^{(4)}(\lambda-1)&= (\lambda^2-1) (\lambda+3) \kappa^{(5)}(\lambda-\frac{1}{2}), \label{psi2}\\
\kappa_{II}^{(4)}(\lambda)\kappa^{(5)}(\lambda-\frac{5}{2})&= (\lambda+3) \kappa_{I}^{(4)}(\lambda-2), 
\nonumber
\end{align}
or, after taking the logarithmic derivative,
\begin{align}
\omega_{II}^{(4)}(\lambda)+\omega_{I}^{(4)}(\lambda-3)&= \frac{1}{\lambda+1}+\frac{1}{\lambda-1}+ \frac{1}{\lambda+3}+\frac{1}{\lambda-3}, \label{omega1} \\
\omega_{II}^{(4)}(\lambda)+\omega_{II}^{(4)}(\lambda-1)&= \frac{1}{\lambda+1}+\frac{1}{\lambda-1}+\frac{1}{\lambda+3} + \omega^{(5)}(\lambda-\frac{1}{2}), \label{omega2}\\
\omega_{II}^{(4)}(\lambda)+\omega^{(5)}(\lambda-\frac{5}{2})&= \frac{1}{\lambda+3}+ \omega_{I}^{(4)}(\lambda-2). \label{omega3}
\end{align}
A simple functional equation can be obtained for $\omega^{(5)}(\lambda)$ by elimination and takes the form
\eq
\omega^{(5)}(\lambda+\tfrac{3}{2})+\omega^{(5)}(\lambda-\tfrac{3}{2})=\frac{1}{\lambda+4}+\frac{1}{\lambda-1}.\label{kappa5FuncEq}
\en
The logarithmic derivatives have the advantage that they admit Fourier-Laplace transforms.

As explained in Appendix B, these inversion relations can be solved in terms of gamma functions or expressed in terms of integrals. 
First the logarithmic derivatives are obtained and then the partition functions per site are obtained by integration.
Observing that the patterns of zeros of $\Lambda_0^{(4)}(\lambda)$ and $\Lambda_0^{(5)}(\lambda)$ in Figure~\ref{fig1} are invariant under the crossing involution $\lambda\mapsto -3-\lambda$ we deduce that
\begin{alignat}{3}
&\mbox{}\hspace{-.1in}\kappa_{I}^{(4)}(\lambda)=\kappa_{II}^{(4)}(-3-\lambda),\quad &&\kappa_{II}^{(4)}(\lambda)=\kappa_{I}^{(4)}(-3-\lambda),\quad &&\kappa^{(5)}(\lambda)=\kappa^{(5)}(-3-\lambda)\label{symmetries1}\\
&\mbox{}\hspace{-.1in}\omega_{I}^{(4)}(\lambda)=-\omega_{II}^{(4)}(-3-\lambda),\quad &&\omega_{II}^{(4)}(\lambda)=-\omega_{I}^{(4)}(-3-\lambda),\quad &&\omega^{(5)}(\lambda)=-\omega^{(5)}(-3-\lambda).\label{symmetries2}
\end{alignat}
Using (\ref{kappa5FuncEq}), we find that the form of (\ref{omega1}) and (\ref{omega3}) are invariant under crossing whereas (\ref{omega2}) belongs to a pair of equations related by crossing. Considering functions of the real variable $\lambda$, the even functions under crossing (\ref{symmetries1}) must be continuous at the crossing point $\lambda=-\tfrac{3}{2}$ whereas the odd functions (\ref{symmetries2}) can exhibit a discontinuity. So some care needs to be taken in solving with two-sided Fourier-Laplace transforms.

Working with gamma functions, we find the result
\bear
\omega_{II}^{(4)}(\lambda)&=&-\frac{1}{3}-\frac{\lambda}{3}\frac{d}{d\lambda}\log{\left[\frac{\Gamma(\frac{1}{2}-\frac{\lambda}{2})\Gamma(\frac{1}{2}+\frac{\lambda}{2})]}{\Gamma(1-\frac{\lambda}{2})\Gamma(1+\frac{\lambda}{2})}\right]} \nonumber \\
&-&\frac{1}{3}\frac{d}{d\lambda}\log{\left[\frac{\Gamma(\frac{1}{3}+\frac{\lambda}{6})\Gamma(\frac{5}{6}-\frac{\lambda}{6})\Gamma(\frac{2}{3}-\frac{\lambda}{6})\Gamma(\frac{1}{6}+\frac{\lambda}{6})}{\Gamma(\frac{1}{3}-\frac{\lambda}{6})\Gamma(\frac{5}{6}+\frac{\lambda}{6})\Gamma(\frac{2}{3}+\frac{\lambda}{6})\Gamma(\frac{1}{6}-\frac{\lambda}{6})}\right]} \label{solsp4} \\
&+&\frac{d}{d\lambda}\log{\left[\frac{\Gamma(1-\frac{\lambda}{6})\Gamma(\frac{3}{2}+\frac{\lambda}{6})\Gamma(\frac{2}{3}-\frac{\lambda}{6})\Gamma(\frac{7}{6}+\frac{\lambda}{6})}{\Gamma(1+\frac{\lambda}{6})\Gamma(\frac{1}{2}-\frac{\lambda}{6})\Gamma(\frac{2}{3}+\frac{\lambda}{6})\Gamma(\frac{1}{6}-\frac{\lambda}{6})}\right]}.  \nonumber
\ear
The last term in (\ref{solsp4}) would be the solution of (\ref{omega1}) if the eigenvalues were analytic in the entire physical strip. Consequently, the other terms in (\ref{solsp4}) can be seen as additional terms resembling CDD factors, although in our case the factors are due to an
extended singularity at the center of the strip \mbox{$-\frac{7}{2}<\mbox{Re}(\lambda)<\frac{1}{2}$}, instead
of an infinite number of isolated singularities.
To the best of our knowledge this is the first instance of a rational model which presents such an intricate structure.

It is worth noticing that (\ref{solsp4}) is the function giving the energy of the quantum chain in the physical strip. This is done by taking the homogeneous limit ($\lambda \to 0$), such that
\bear
\omega_{II}^{(4)}(0)=1-\frac{2\pi}{9\sqrt3}- \frac{4}{3} \log2,
\label{groundstate}
\ear
which is, apart from a trivial shift of the whole spectrum due to different normalization, the ground state energy of the quantum spin chain studied via the solution of the Bethe ansatz equations in \cite{MARTINS-SP2N}.

Working in terms of gamma functions, the solution of the remaining functions $\omega_{I}^{(4)}(\lambda-2)$ and $\omega^{(5)}(\lambda-\tfrac{1}{2})$ are given by
\bear
\omega_{I}^{(4)}(\lambda-2)&=&
-\frac{1}{3}-\frac{\lambda}{3}\frac{d}{d\lambda}\log{\left[\frac{\Gamma(\frac{1}{2}-\frac{\lambda}{2})\Gamma(\frac{1}{2}+\frac{\lambda}{2})]}{\Gamma(1-\frac{\lambda}{2})\Gamma(1+\frac{\lambda}{2})}\right]} \nonumber \\
&-&\frac{1}{3}\frac{d}{d\lambda}\log{\left[\frac{\Gamma(\frac{1}{3}+\frac{\lambda}{6})\Gamma(\frac{5}{6}-\frac{\lambda}{6})\Gamma(\frac{2}{3}-\frac{\lambda}{6})\Gamma(\frac{1}{6}+\frac{\lambda}{6})}{\Gamma(\frac{1}{3}-\frac{\lambda}{6})\Gamma(\frac{5}{6}+\frac{\lambda}{6})\Gamma(\frac{2}{3}+\frac{\lambda}{6})\Gamma(\frac{1}{6}-\frac{\lambda}{6})}\right]} \label{solwI4} \\
&+&\frac{d}{d\lambda}\log{\left[\frac{\Gamma(1-\frac{\lambda}{6})\Gamma(\frac{4}{3}-\frac{\lambda}{6})\Gamma(\frac{2}{3}-\frac{\lambda}{6})\Gamma(\frac{7}{6}+\frac{\lambda}{6})}{\Gamma(\frac{1}{2}-\frac{\lambda}{6})\Gamma(\frac{1}{6}-\frac{\lambda}{6})\Gamma(\frac{2}{3}+\frac{\lambda}{6})\Gamma(\frac{5}{6}-\frac{\lambda}{6})} \right]}, \nonumber
\ear 
\bear 
\omega^{(5)}(\lambda-\frac{1}{2})&=&\frac{d}{d\lambda}\log{\left[\frac{\Gamma(1-\frac{\lambda}{6})\Gamma(\frac{4}{3}+\frac{\lambda}{6})}{\Gamma(\frac{1}{2}-\frac{\lambda}{6})\Gamma(\frac{5}{6}+\frac{\lambda}{6})} \right]}. \label{solw5}
\ear

By integrating (\ref{solsp4}) and fixing the integration constants such that the unitarity property is satisfied, we obtain the partition function per site given by,
\bear
\kappa_{II}^{(4)}(\lambda)&=&6^2e^{-\frac{\lambda}{3}}e^{\frac{2}{3} h(\lambda)} 
\left[\frac{\Gamma(1-\frac{\lambda}{6})\Gamma(\frac{3}{2}+\frac{\lambda}{6})\Gamma(\frac{2}{3}-\frac{\lambda}{6})\Gamma(\frac{7}{6}+\frac{\lambda}{6})}{\Gamma(1+\frac{\lambda}{6})\Gamma(\frac{1}{2}-\frac{\lambda}{6})\Gamma(\frac{2}{3}+\frac{\lambda}{6})\Gamma(\frac{1}{6}-\frac{\lambda}{6})}\right] \nonumber \\
&\times&  \left[\frac{\Gamma(\frac{1}{3}-\frac{\lambda}{6})\Gamma(\frac{5}{6}+\frac{\lambda}{6})\Gamma(\frac{2}{3}+\frac{\lambda}{6})\Gamma(\frac{1}{6}-\frac{\lambda}{6})}{\Gamma(\frac{1}{3}+\frac{\lambda}{6})\Gamma(\frac{5}{6}-\frac{\lambda}{6})\Gamma(\frac{2}{3}-\frac{\lambda}{6})\Gamma(\frac{1}{6}+\frac{\lambda}{6})}\right]^{\frac{1}{3}}\left[\frac{\Gamma(1-\frac{\lambda}{2})\Gamma(1+\frac{\lambda}{2})}{\Gamma(\frac{1}{2}-\frac{\lambda}{2})\Gamma(\frac{1}{2}+\frac{\lambda}{2})}\right]^{\frac{\lambda}{3}},
\ear
where
\eq
h(\lambda)=\psi^{(-2)}(1-\frac{\lambda}{2})-\psi^{(-2)}(1+\frac{\lambda}{2})+\psi^{(-2)}(\frac{1}{2}+\frac{\lambda}{2})-\psi^{(-2)}(\frac{1}{2}-\frac{\lambda}{2}),
\en
and the function $\psi^{(-2)}(\lambda)$ is defined as \cite{POLYNEG},
\eq
\psi^{(-2)}(\lambda)=\int_0^{\lambda}  \log(\Gamma(t)) dt.
\en

\begin{figure}[th]
	\begin{center}
		\includegraphics[width=0.65\linewidth, angle =-90 ]{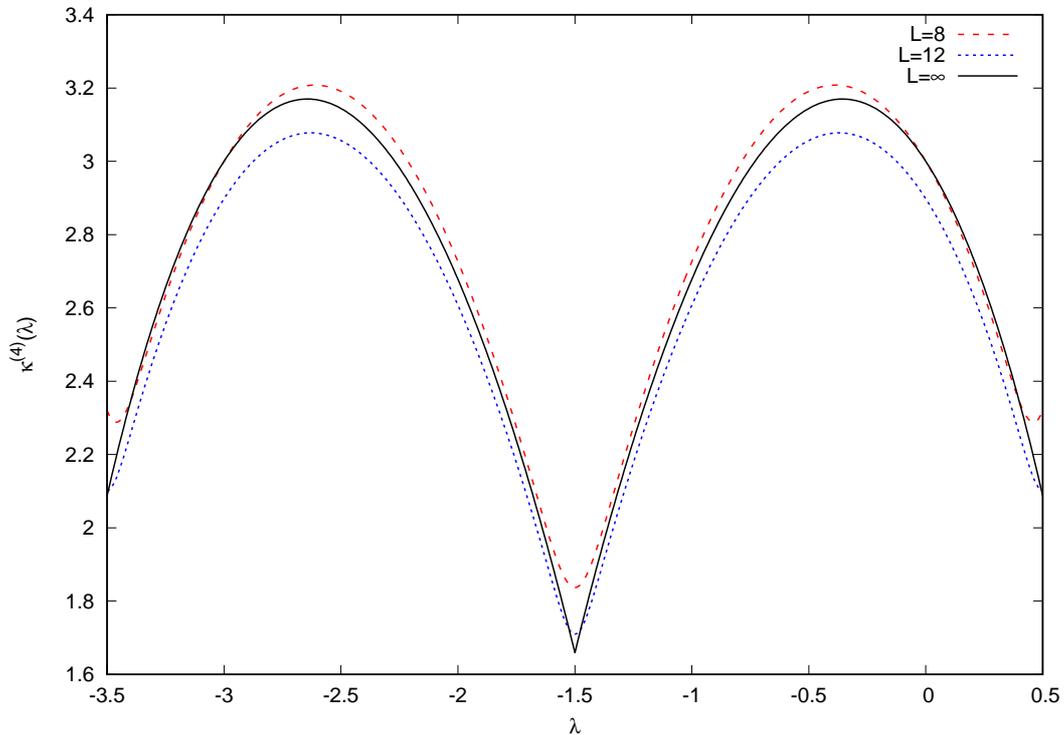}
		\caption{Partition function per site $\kappa^{(4)}_L(\lambda)=\left(\Lambda^{(4)}_0(\lambda)\right)^{1/L}$ as a function of $\lambda$ for finite horizontal lattice sizes $L=8,12$ and its comparison with the result in the thermodynamic limit ($L\rightarrow\infty$). Notice that $\kappa^{(4)}_L(\lambda)$ is analytic for $L$ finite but develops a cusp and is not analytic at $\lambda=-\tfrac{3}{2}$ for $L\to\infty$.}
		\label{fig2}
	\end{center}
\end{figure}

Again the remaining functions $\kappa_{I}^{(4)}(\lambda-2)$ and $\kappa^{(5)}(\lambda-\frac{1}{2})$ can be  obtained by direct integration of (\ref{solwI4}--\ref{solw5}), whose results are listed below.
\bear
\kappa_{I}^{(4)}(\lambda-2)&=& 6^2 e^{-\frac{\lambda}{3}}e^{\frac{2}{3} h(\lambda)} 
\left[\frac{\Gamma(1-\frac{\lambda}{6})\Gamma(\frac{4}{3}-\frac{\lambda}{6})\Gamma(\frac{2}{3}-\frac{\lambda}{6})\Gamma(\frac{7}{6}+\frac{\lambda}{6})}{\Gamma(\frac{1}{2}-\frac{\lambda}{6})\Gamma(\frac{1}{6}-\frac{\lambda}{6})\Gamma(\frac{2}{3}+\frac{\lambda}{6})\Gamma(\frac{5}{6}-\frac{\lambda}{6})}\right]  \\
&\times&  \left[\frac{\Gamma(\frac{1}{3}-\frac{\lambda}{6})\Gamma(\frac{5}{6}+\frac{\lambda}{6})\Gamma(\frac{2}{3}+\frac{\lambda}{6})\Gamma(\frac{1}{6}-\frac{\lambda}{6})}{\Gamma(\frac{1}{3}+\frac{\lambda}{6})\Gamma(\frac{5}{6}-\frac{\lambda}{6})\Gamma(\frac{2}{3}-\frac{\lambda}{6})\Gamma(\frac{1}{6}+\frac{\lambda}{6})}\right]^{\frac{1}{3}}\left[\frac{\Gamma(1-\frac{\lambda}{2})\Gamma(1+\frac{\lambda}{2})}{\Gamma(\frac{1}{2}-\frac{\lambda}{2})\Gamma(\frac{1}{2}+\frac{\lambda}{2})}\right]^{\frac{\lambda}{3}},\nonumber\\
\kappa^{(5)}(\lambda-\frac{1}{2})&=& 6 \left[ \frac{\Gamma(1-\frac{\lambda}{6})\Gamma(\frac{4}{3}+\frac{\lambda}{6})}{\Gamma(\frac{1}{2}-\frac{\lambda}{6})\Gamma(\frac{5}{6}+\frac{\lambda}{6})}\right]. 
\ear

\begin{figure}[th]
	\begin{center}
		\includegraphics[width=0.65\linewidth, angle=-90]{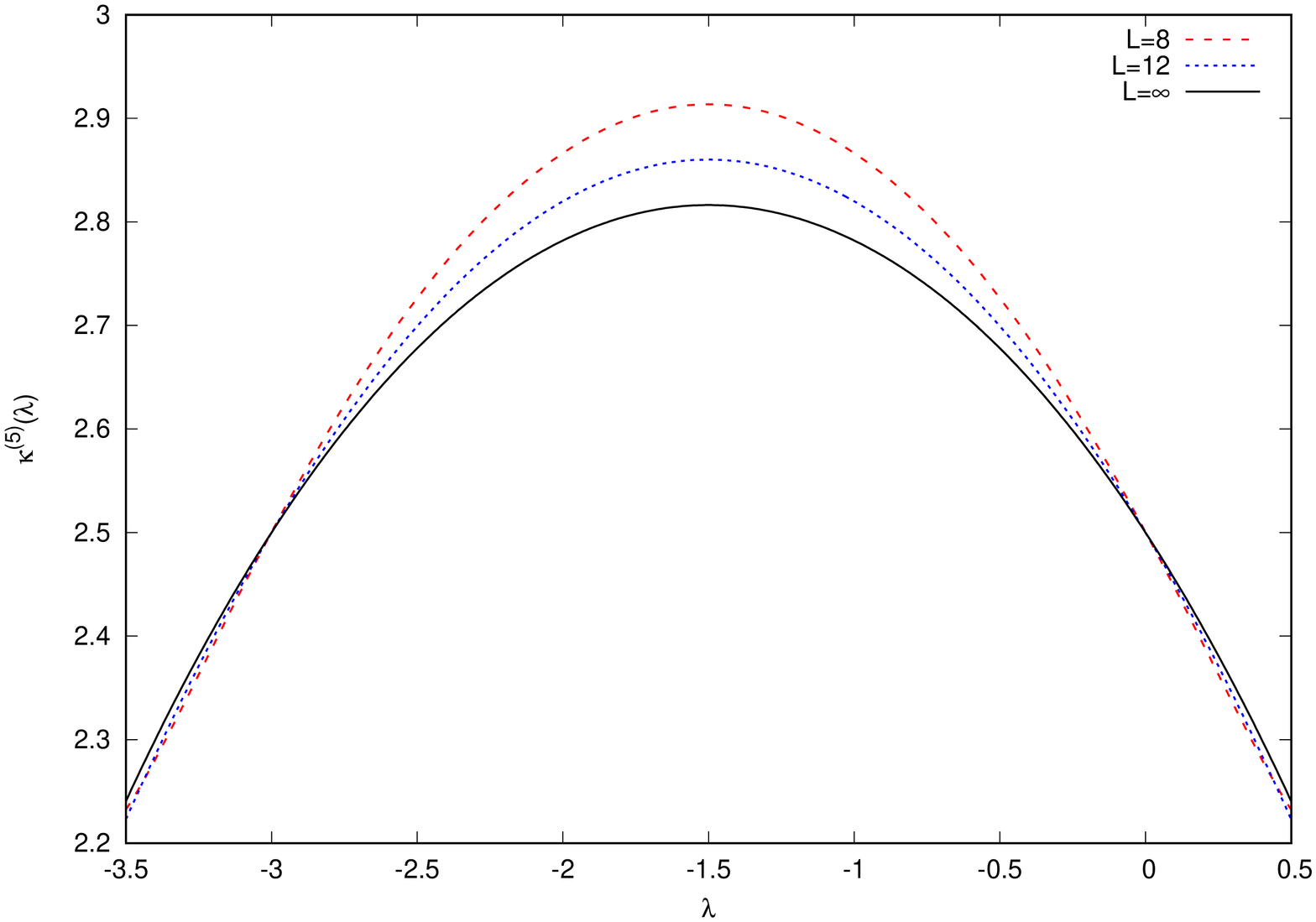}
		\caption{Partition function per site $\kappa^{(5)}_L(\lambda)=\left(\Lambda^{(5)}_0(\lambda)\right)^{1/L}$ as a function of $\lambda$ for finite horizontal lattice sizes $L=8,12$ and its comparison with the result in the thermodynamic limit ($L\rightarrow\infty$).}
		\label{fig3}
	\end{center}
\end{figure}

In order to further illustrate, we show in Figures \ref{fig2} and \ref{fig3} the comparison of the partition functions per site  $\kappa^{(\alpha)}_L(\lambda)=\left(\Lambda^{(\alpha)}_0(\lambda)\right)^{1/L}$ for $\alpha=4,5$ for finite horizontal lattice sizes $L=8,12$ as a function of $\lambda$ with the result in the thermodynamic limit.

\section{Conclusion}\label{CONCLUSION}

We investigated the partition function of the fundamental $Sp(4)$ vertex model on a square lattice in the thermodynamic limit via an approach based on functional equations. 

This is a subtle problem due to the special analytic properties of the partition function. 
This is due to the fact that the leading eigenvalue of the transfer matrix has zeros in the complex plane distributed along a line at the center of the physical strip. Besides that, the number of these zeros grows linearly with the system size. Therefore, in the thermodynamic limit, the partition function per site is no longer an analytic function across this line. In order to fix this, we exploit a suitable set of functional relations for the transfer matrix obtained by fusion.

We solve the resulting relations in the thermodynamic limit, which fully determine the partition function of the model. The result is given in terms of the solution based on the assumption of analyticity in the full physical strip plus additional terms which can be seen as a kind of generalized CDD factor, which in our case is produced by an extended singularity instead of isolated singularities. In case of usual CDD factors the ambiguities in analytic continuations around
the singularities are multiples of $2\pi \im$. In our case the jump function is	rather non-trivial, see (\ref{jump}). In addition, as a byproduct, we also investigated the partition function of a vertex model  mixing the four and five dimensional representations of the $Sp(4)$ symmetry. 

Besides the results for the vertex model on the square lattice, the determin\-ation of the partition function is an essential first step toward the calculation of the correlation functions via quantum Knizhnik-Zamolodchikov equations. This is due to the fact that the function $\omega_{i}^{(4)}(\lambda)$ is precisely the first of two functions 
needed to fully determine the two-sites correlation of the $Sp(4)$ quantum spin chain. Nevertheless, in this context one still needs to formulate fused quantum Knizhnik-Zamolodchikov equations to obtain all the additional fused functional equations. We also expect that the method used in this work can be extended to the general case of $Sp(2n)$ as well as to other models with similar analytic subtleties. We hope to address theses problems in the future.

\section*{Acknowledgments}

GAPR thanks for support and hospitality the University of Melbourne where this work started and the University of Wuppertal where this work was essentially completed. GAPR acknowledges the S\~ao Paulo Research Foundation (FAPESP) for financial support through the grant 2018/25824-0. GAPR also thanks M.J. Martins for discussions and careful reading of the manuscript.

\section*{Appendix A: $Sp(4)$ projectors and fusion rules}

In this appendix, we present more details concerning fusion in the $Sp(4)$ case.

The projectors $\check{P}_{12}^{(\alpha)}$ for $\alpha=1,5,10$ which arise from the decomposition $4\otimes 4= 1\oplus 5 \oplus 10$ are simply related to the identity, permutation and Temperley-Lieb operators already defined. Therefore, we just list their explicit relations as follows,
\begin{align}
\check{P}_{12}^{(1)}&=-\frac{1}{4} E_{12}, \\
\check{P}_{12}^{(5)}&= \frac{1}{2} \left( I_{12} -  P_{12} \right) + \frac{1}{4} E_{12}, \\
\check{P}_{12}^{(10)}&= \frac{1}{2}\left( I_{12} +  P_{12} \right). 
\end{align}

The occurrence of the projectors $\check{P}_{12}^{(\alpha)}$ for $\alpha=4,16$ is due to the decomposition $5\otimes 4= 4\oplus 16$~\cite{GROUP}. Since $\check{P}_{12}^{(4)}+\check{P}_{12}^{(16)}=I$, we only list the projector on the $4$-dimensional space, 
\bear
\footnotesize
\arraycolsep=1.5pt
\check{P}_{12}^{(4)}\!=\!
\frac{1}{5}\!
\left(\begin{array}{cccc|cccc|cccc|cccc|cccc}
    0 & 0 & 0 & 0 & 0 & 0 & 0 & 0 & 0 & 0 & 0 & 0 & 0 & 0 & 0 & 0 & 0 & 0 & 0 & 0\\[-8pt]
	0 & 0 & 0 & 0 & 0 & 0 & 0 & 0 & 0 & 0 & 0 & 0 & 0 & 0 & 0 & 0 & 0 & 0 & 0 & 0\\[-8pt] 
	0 & 0 & 2 & 0 & 0 & -2 & 0 & 0 & -\sqrt{2} & 0 & 0 & 0 & 0 & 0 & 0 & 0 & 0 & 0 & 0 & 0\\[-8pt] 
		0 & 0 & 0 & 2 & 0 & 0 & 0 & 0 & 0 & -\sqrt{2} & 0 & 0 & 2 & 0 & 0 & 0 & 0 & 0 & 0 & 0\\[-2pt]
\hline\\[-20pt]		 
		0 & 0 & 0 & 0 & 0 & 0 & 0 & 0 & 0 & 0 & 0 & 0 & 0 & 0 & 0 & 0 & 0 & 0 & 0 & 0\\[-8pt] 
	0 & 0 & -2 & 0 & 0 & 2 & 0 & 0 & \sqrt{2} & 0 & 0 & 0 & 0 & 0 & 0 & 0 & 0 & 0 & 0 & 0\\[-8pt]
		0 & 0 & 0 & 0 & 0 & 0 & 0 & 0 & 0 & 0 & 0 & 0 & 0 & 0 & 0 & 0 & 0 & 0 & 0 & 0\\[-8pt]
	0 & 0 & 0 & 0 & 0 & 0 & 0 & 2 & 0 & 0 & -\sqrt{2} & 0 & 0 & 0 & 0 & 0 & 2 & 0 & 0 & 0\\[-2pt]
\hline\\[-20pt]	
		0 & 0 & -\sqrt{2} & 0 & 0 & \sqrt{2} & 0 & 0 & 1 & 0 & 0 & 0 & 0 & 0 & 0 & 0 & 0 & 0 & 0 & 0\\[-8pt]
		0 & 0 & 0 & -\sqrt{2} & 0 & 0 & 0 & 0 & 0 & 1 & 0 & 0 & -\sqrt{2} & 0 & 0 & 0 & 0 & 0 & 0 & 0\\[-8pt]
		0 & 0 & 0 & 0 & 0 & 0 & 0 & -\sqrt{2} & 0 & 0 & 1 & 0 & 0 & 0 & 0 & 0 & -\sqrt{2} & 0 & 0 & 0\\[-8pt]
		0 & 0 & 0 & 0 & 0 & 0 & 0 & 0 & 0 & 0 & 0 & 1 & 0 & 0 & \sqrt{2} & 0 & 0 & -\sqrt{2} & 0 & 0\\[-2pt]
\hline\\[-20pt]		 
	0 & 0 & 0 & 2 & 0 & 0 & 0 & 0 & 0 & -\sqrt{2} & 0 & 0 & 2 & 0 & 0 & 0 & 0 & 0 & 0 & 0\\[-8pt] 
		0 & 0 & 0 & 0 & 0 & 0 & 0 & 0 & 0 & 0 & 0 & 0 & 0 & 0 & 0 & 0 & 0 & 0 & 0 & 0\\[-8pt] 
	0 & 0 & 0 & 0 & 0 & 0 & 0 & 0 & 0 & 0 & 0 & \sqrt{2} & 0 & 0 & 2 & 0 & 0 & -2 & 0 & 0\\[-8pt]	 
		0 & 0 & 0 & 0 & 0 & 0 & 0 & 0 & 0 & 0 & 0 & 0 & 0 & 0 & 0 & 0 & 0 & 0 & 0 & 0\\[-2pt] 
\hline\\[-20pt]		
	0 & 0 & 0 & 0 & 0 & 0 & 0 & 2 & 0 & 0 & -\sqrt{2} & 0 & 0 & 0 & 0 & 0 & 2 & 0 & 0 & 0\\[-8pt] 
		0 & 0 & 0 & 0 & 0 & 0 & 0 & 0 & 0 & 0 & 0 & -\sqrt{2} & 0 & 0 & -2 & 0 & 0 & 2 & 0 & 0\\[-8pt] 
		0 & 0 & 0 & 0 & 0 & 0 & 0 & 0 & 0 & 0 & 0 & 0 & 0 & 0 & 0 & 0 & 0 & 0 & 0 & 0\\[-8pt] 
		0 & 0 & 0 & 0 & 0 & 0 & 0 & 0 & 0 & 0 & 0 & 0 & 0 & 0 & 0 & 0 & 0 & 0 & 0 & 0\\[-2pt]
\end{array}\right)\!. 
\ear

\normalsize

We now list the main fusion rules needed in this work \cite{CAO}. 
By exploiting the singular values of $R_{12}^{(4,4)}(\lambda)$ we have that,
\begin{align}
\check{P}_{ab}^{(1)} R_{b2}^{(4,4)}(\lambda)R_{a2}^{(4,4)}(\lambda-3)\check{P}_{ab}^{(1)} &=(\lambda^2-1)(\lambda^2-3^2)\check{P}_{ab}^{(1)},  \label{fus1} \\
\check{P}_{ab}^{(5)}R_{b2}^{(4,4)}(\lambda)R_{a2}^{(4,4)}(\lambda-1)\check{P}_{ab}^{(5)}&=(\lambda^2-1)(\lambda+3) ~ R_{12}^{(5,4)}(\lambda-\frac{1}{2}), \label{fus2}
\end{align}
where we recall that the dimension of each space is indicated by the upper index, which means that the first space in $R_{12}^{(5,4)}(\lambda)$ is the $5$-dimensional representation. 

Again, we can use the singular point of $R_{12}^{(5,4)}(\lambda)$ to obtain,
\eq
\check{P}_{ab}^{(4)} R_{b2}^{(4,4)}(\lambda)R_{a2}^{(5,4)}(\lambda-\frac{5}{2})\check{P}_{ab}^{(4)}=(\lambda+3) R_{12}^{(4,4)}(\lambda-2). \label{fus3}
\en

Similar relations exist for the product of monodromy matrices, which are ordered products of the $R$-matrices ${\cal T}_{\cal A}^{(\alpha,4)}(\lambda)=R_{{\cal A}L}^{(\alpha,4)}(\lambda)\cdots R_{{\cal A}1}^{(\alpha,4)}(\lambda)$, with \mbox{$\alpha=4,5$}. Therefore, the transfer matrix fusion relations (\ref{TMFR1}-\ref{TMFR3}) are naturally obtained from the fusion relations. 

For instance, by inserting the identity as the sum of the projectors into the trace, moving them around the trace and finally by using (\ref{fus1}) we see that,
\bear
&&T^{(4)}(\lambda)T^{(4)}(\lambda-3)=\tr_{a \otimes b}{\left[ {\cal T}_{b}^{(4,4)}(\lambda) {\cal T}_{a}^{(4,4)}(\lambda-3)\right]}, \nonumber \\
&=& \tr_{a \otimes b}{\left[ \left(P_{ab}^{(1)} + P_{ab}^{(5)} +P_{ab}^{(10)}  \right) {\cal T}_{b}^{(4,4)}(\lambda) {\cal T}_{a}^{(4,4)}(\lambda-3)\right]}, \\
&=& \tr_{a \otimes b}{\left[ P_{ab}^{(1)}  {\cal T}_{b}^{(4,4)}(\lambda) {\cal T}_{a}^{(4,4)}(\lambda-3) P_{ab}^{(1)} \right]} \nonumber\\
&+& \sum_{\alpha=5,10}\tr_{a \otimes b}{\left[ P_{ab}^{(\alpha)}  {\cal T}_{b}^{(4,4)}(\lambda) {\cal T}_{a}^{(4,4)}(\lambda-3) P_{ab}^{(\alpha)} \right]}, \nonumber \\
&=& \left[(\lambda^2-1)(\lambda^2-3^2)\right]^L I \nonumber\\
&+& \sum_{\alpha=5,10}\tr_{a \otimes b}{\left[ P_{ab}^{(\alpha)}  {\cal T}_{b}^{(4,4)}(\lambda) {\cal T}_{a}^{(4,4)}(\lambda-3) P_{ab}^{(\alpha)} \right]}, \nonumber
\ear
which gives the transfer matrix inversion identity (\ref{TMFI1}) provided that the auxiliary matrix $V_1(\lambda)$ is defined as,
\eq
\varphi(\lambda)V_1(\lambda)=\sum_{\alpha=5,10}\tr_{a \otimes b}{\left[ P_{ab}^{(\alpha)}  {\cal T}_{b}^{(4,4)}(\lambda) {\cal T}_{a}^{(4,4)}(\lambda-3) P_{ab}^{(\alpha)} \right]}.
\label{auxV1}
\en 
It is worth noticing that (\ref{auxV1}) is trivially zero at $\lambda=0$, due to the product of projection operators on different subspaces.

The remaining transfer matrix fusion identities (\ref{TMFR2}--\ref{TMFR3}) are obtained along the same lines as above.

\section*{Appendix B: Solution of inversion relations}

The functional equations (\ref{omega1}--\ref{omega3}) are solved by Fourier-Laplace transform. This means that the equations are Fourier-Laplace transformed and afterwards the system of the resulting equation is solved algebraically. Finally, we perform the inverse transformation which gives the full solution in integral form. We conveniently write the integrals in terms of digamma functions using the following identity,    
\eq
\psi(\lambda)=\frac{d}{d\lambda}\log\Gamma(\lambda)=-\gamma-\int_0^\infty \frac{1-e^{(1-\lambda) t}}{1-e^t}dt\;\sim\; -\int_0^\infty \frac{e^{-\lambda t}}{1-e^{-t}} \,dt
\en
where $\psi(\lambda)$ is the digamma function, $\gamma$ is Euler's constant and only the last integral is relevant here since only differences of digamma functions occur. The partition functions per site $\kappa(\lambda)$ are subsequently obtained by integration, whose results are naturally given in terms of $\Gamma$-functions. Alternatively, we can also write the results closer to the spirit of the method of Baxter \cite{BAXTER-inversion}, which instead presents the result as an integral expression for $\log\kappa(\lambda)$.

For simplicity in illustrating this, 
let us consider only the case of  $\omega^{(5)}(\lambda)$ and $\kappa^{(5)}(\lambda)$ which satisfy the functional equations
\eq
\omega^{(5)}(\lambda+\tfrac{3}{2})+\omega^{(5)}(\lambda-\tfrac{3}{2})=\frac{1}{\lambda+4}+\frac{1}{\lambda-1},\qquad 
\kappa^{(5)}(\lambda)\kappa^{(5)}(-\lambda)=(\tfrac{5}{2})^2-\lambda^2.
\en
The first functional equation follows by elimination in (\ref{omega1}--\ref{omega3}) and the second is unitarity. 

The solution to the above functional equation can be obtained e.g. by two-sided Laplace transform. Since the Laplace transform  for $\omega^{(5)}(\lambda)$ is additive, let us first consider the contribution from the term $\frac{1}{\lambda+4}$. Writing $\omega^{(5)}(\lambda)$ as a Laplace integral
\eq
\omega^{(5)}(\lambda)={\cal L}\{a(t)\}=\int_0^\infty a(t) e^{-\lambda t}\,dt,
\en
it follows that
\eq
\omega^{(5)}(\lambda-\tfrac{3}{2})+\omega^{(5)}(\lambda+\tfrac{3}{2})=2\!\int_0^\infty\!\!a(t) \cosh\tfrac{3t}{2}\, e^{-\lambda t}\,dt.
\en
Taking inverse Laplace transforms gives
\eq
2\,a(t) \cosh\tfrac{3t}{2}={\cal L}^{-1}\{\tfrac{1}{\lambda+4}\}=e^{-4t},\quad a(t)=\frac{e^{-4t}}{2\cosh\tfrac{3t}{2}},\quad
\omega^{(5)}(\lambda)=\int_0^\infty \frac{e^{-(\lambda+4)t}}{2\cosh\tfrac{3t}{2}}\,dt.
\en
To get the other contribution, arising from $\frac{1}{\lambda-1}$, we need to work on the $t<0$ half-line with $e^{\lambda t}$ in the Laplace transforms. Putting these together is equivalent to working from the outset with two-sided Laplace transforms and treating $t>0$ and $t<0$ separately.
After some simplification, this yields the integral representation
\eq
\omega^{(5)}(\lambda)=-\int_0^\infty \!\! e^{-\frac{5t}{2}}\, \frac{\sinh(\lambda+\frac{3}{2}) t}{\cosh{\frac{3t}{2}}}\,dt=\int_0^\infty \frac{e^{-(\lambda+4)t}-e^{(\lambda-1)t}}{2\cosh{\frac{3t}{2}}}\,dt.\label{IntOmega5}
\en
This integral can be recast in terms of the digamma function as follows,
\begin{align}
\omega^{(5)}(\lambda)&=\int_0^\infty  \frac{e^{-(\lambda+\frac{11}{2})t}-e^{(\lambda-\frac{5}{2})t}}{1+e^{-3t}}\,dt\;=\int_0^\infty  \frac{[e^{-(\lambda+\frac{11}{2})t}-e^{(\lambda-\frac{5}{2})t}]({1-e^{-3t}})}{1-e^{-6t}}\,dt, \nonumber\\
&=\int_0^\infty  \left[\frac{e^{-(\frac{11}{2}+\lambda)t}}{1-e^{-6t}}- \frac{e^{-(\frac{17}{2}+\lambda)t}}{1-e^{-6t}}\right]dt -\int_0^\infty  \left[\frac{e^{-(\frac{5}{2}-\lambda)t}}{1-e^{-6t}}- \frac{e^{-(\frac{11}{2}-\lambda)t}}{1-e^{-6t}}\right]dt, \nonumber\\
&=\tfrac{1}{6}[ -\psi(\tfrac{11}{12}+\tfrac{\lambda}{6}) +\psi(\tfrac{17}{12}+\tfrac{\lambda}{6})   +\psi(\tfrac{5}{12}-\tfrac{\lambda}{6}) -\psi(\tfrac{11}{12}-\tfrac{\lambda}{6})], \\
&=\frac{d}{d\lambda}\log\left[\frac{\Gamma(\frac{17}{12}+\frac{\lambda}{6}) \Gamma(\frac{11}{12}-\frac{\lambda}{6})}{\Gamma(\frac{11}{12}+\frac{\lambda}{6}) \Gamma(\frac{5}{12}-\frac{\lambda}{6})} \right], \nonumber
\end{align}
in agreement with (\ref{solw5}).
Integrating (\ref{IntOmega5}), using the initial value $\kappa^{(5)}(0)=\tfrac{5}{2}$ and simplifying gives
\eq
\log \kappa^{(5)}(\lambda)=\log\tfrac{5}{2}-2\!\int_0^\infty e^{-\frac{5t}{2}}\,\frac{\sinh \frac{\lambda t}{2}\sinh \frac{(3+\lambda)t}{2}}{t \cosh \frac{3t}{2}}\,dt.
\en

Integral expressions in the other regions are obtained similarly giving the integral representations
\begin{align}
	\omega^{(4)}_I(\lambda)&=-\int_0^\infty  \frac{\cosh(\lambda+3)t}{2\cosh\frac{t}{2}\cosh\frac{3t}{2}}\,dt - 2 \int_0^\infty e^{-2 t}\, \frac{\cosh t \sinh(\lambda+\frac{3}{2})t}{\cosh\frac{3t}{2}}\,dt, 
	\\
	\log\kappa^{(4)}_I(\lambda)&=
	\log3 -\!\!\int_0^\infty  \!\frac{\sinh(\lambda+3) t}{2 t\cosh\frac{t}{2}\cosh\frac{3t}{2}}\,dt - 4\! \int_0^\infty \!\! e^{-2 t}\,\frac{\cosh t\sinh\frac{\lambda t}{2} \sinh\frac{(\lambda+3)t}{2}}{t \cosh\frac{3t}{2}}\,dt,
\end{align}
\begin{align}
	\omega^{(4)}_{II}(\lambda)&=\int_0^\infty  \frac{\cosh\lambda t}{2\cosh\frac{t}{2}\cosh\frac{3t}{2}}\,dt - 2 \int_0^\infty\! e^{-2 t}\, \frac{\cosh t \sinh(\lambda+\frac{3}{2})t}{\cosh\frac{3t}{2}}\,dt, \\
	\log\kappa^{(4)}_{II}(\lambda)&=\log3 +\!\!\int_0^\infty \! \frac{\sinh\lambda t}{2 t\cosh\frac{t}{2}\cosh\frac{3t}{2}}\,dt - 
	4\! \int_0^\infty \!\! e^{-2 t}\,\frac{\cosh t\sinh\frac{\lambda t}{2} \sinh\frac{(\lambda+3)t}{2}}{t \cosh\frac{3t}{2}}\,dt.
\end{align}
It is readily verified that all these functions, expressed as integrals or in terms of gamma functions, display the correct crossing symmetries (\ref{symmetries1}--\ref{symmetries2}). To express the first integral in (55) in terms of digamma functions requires some extra steps involving partial fractions and integration by parts. Doing this, we observe that
\begin{align}
&\qquad\qquad\omega^{(4)}_{II}(\lambda)=-\tfrac{\lambda}{6}[\psi(\tfrac{\lambda+1}{2})\!-\!\psi(\tfrac{\lambda}{2})\!+\!\psi(\tfrac{2-\lambda}{2})\!-\!\psi(\tfrac{1-\lambda}{2})]+\mbox{}\nonumber\\
&\tfrac{1}{18}[\psi(\tfrac{4-\lambda}{6})\!+\!\psi(\tfrac{5-\lambda}{6})\!+\!\psi(\tfrac{\lambda+4}{6})\!+\!\psi(\tfrac{\lambda+5}{6})\nonumber
\!-\!\psi(\tfrac{1-\lambda}{6})\!-\!\psi(\tfrac{\lambda+1}{6})\!-\!\psi(\tfrac{2-\lambda}{6})\!-\!\psi(\tfrac{\lambda+2}{6})]+\mbox{}\nonumber\\
&\tfrac{1}{6}[\psi(\tfrac{1-\lambda}{6})\!+\!\psi(\tfrac{3-\lambda}{6})\!+\!\psi(\tfrac{\lambda+7}{6})\!+\!\psi(\tfrac{\lambda+9}{6})
\!-\!\psi(\tfrac{4-\lambda}{6})\!-\!\psi(\tfrac{\lambda+4}{6})\!-\!\psi(\tfrac{6-\lambda}{6})\!-\!\psi(\tfrac{\lambda+6}{6})]
\end{align}
\eq
\omega^{(4)}_{II}(\lambda)-\omega^{(4)}_{I}(\lambda)=\!\!\int_0^\infty \frac{\cosh(\lambda\!+\!\frac{3}{2})t}{\cosh\frac{t}{2}}\,dt
=\tfrac{1}{2}[\psi(\tfrac{\lambda+1}{2})-\psi(\tfrac{\lambda}{2})+\psi(\tfrac{2-\lambda}{2})-\psi(\tfrac{1-\lambda}{2})]=\frac{\pi}{\sin \pi \lambda}. \label{jump}
\en
It follows that, whereas $\omega_I^{(4)}(\lambda)$ and $\omega_{II}^{(4)}(\lambda)$ can be analytically continued across the cut at the crossing point $\lambda=-\tfrac{3}{2}$, $\omega^{(4)}(\lambda)$ given by (\ref{kappaomega}) exhibits a jump discontinuity of $\pi$.

\renewcommand{\baselinestretch}{1.5}


\begin{thebibliography}{100}
\bibitem{BAXTER} R.J. Baxter \textit{Exactly solved models in statistical mechanics} (AP, London, 1982).
\bibitem{BOOK-KBI} V.E. Korepin, N.M. Bogoliubov, A.G. Izergin \textit{Quantum inverse scattering method and correlation functions} (CUP, Cambridge, 1993).
\bibitem{STROGANOV} Y.G. Stroganov, Phys. Lett. 74A (1979) 116.
\bibitem{BAXTER-inversion} R.J. Baxter, J. Stat. Phys. {28} (1982) 1--41.
\bibitem{SHANKAR} R. Shankar, Phys. Rev. Lett., 47 (1981) 1177.
\bibitem{BAXTER-PEARCE} R.J. Baxter and P.A. Pearce, J. Phys. A: Math. Gen., 15 (1982) 897; 16 (1983) 2239.
\bibitem{KIRILLOV} A.N. Kirillov and N.Yu. Reshetikhin, J. Sov. Math., 35 (1986) 2627; J. Phys. A,
20 (1987) 1565.
\bibitem{BAZHANOV} V.V. Bazhanov and Yu.N. Reshetikhin, Int. J. Mod. Phys. A4 (1989) 115.
\bibitem{PEARCE1987} P.A. Pearce, Phys. Rev. Lett. 58 (1987) 1502.
\bibitem{SPIN-ORBITAL} D.P. Arovas, A. Auerbach, Phys. Rev. B, 52 (1995) 10114; E. Orignac, R. Citro, N. Andrei, Phys. Rev. B, 61 (2000) 11533; A.K. Kolezhuk, H.J. Mikeska, U. Schollwock, Phys. Rev. B, 6305 (2001) 4418.
\bibitem{KAROWSKI} M. Karowski, Nucl. Phys. B, 153 (1979) 244.
\bibitem{KULISH1981} P. P. Kulish, N. Yu. Reshetikhin and E. K. Sklyanin, Lett. Math. Phys. 5 (1981) 393.
\bibitem{KULISH1982} P. P. Kulish and E. K. Sklyanin, Lecture Notes in Physics, 151 (1982) 61.
\bibitem{CAO} G.-L. Li, J. Cao, P. Xue, Z.-R. Xin, K. Hao, W.-L. Yang, K. Shi, Y. Wang, JHEP, 05 (2019) 067.
\bibitem{RIBEIRO} G. A. P. Ribeiro, Nucl. Phys. B, 957 (2020) 115106.
\bibitem{BOOS05} H.~Boos, M.~Jimbo, T.~Miwa, F.~Smirnov and Y.~Takeyama, Algebra and Analysis, 17 (2005) 115; H.~Boos, M.~Jimbo, T.~Miwa, F.~Smirnov and Y.~Takeyama, St. Petersburg Math. J.,  17 (2006) 85. 
\bibitem{BOOS2}  H. Boos, M. Jimbo, T. Miwa, F. Smirnov, Y. Takeyama, Comm. Math. Phys. 261 (2006) 245.
\bibitem{RESHETIKHIN} N. Yu. Reshetikhin, Lett. Math. Phys., 7 (1983) 205; N. Yu. Reshetikhin,  Theor. Math. Fiz. 63 (1985) 347;  N. Yu. Reshetikhin, Lett. Math. Phys, 14 (1987) 235.
\bibitem{KUNIBA} A. Kuniba and J. Suzuki, Comm.Math.Phys. 173 (1995) 225.
\bibitem{KULISH} P.P. Kulish, J. Sov. Math., 35 (1986) 2648.
\bibitem{MARTINS1997} M. J. Martins, P. B. Ramos, Nucl. Phys. B, 500 (1997) 579.
\bibitem{MARTINS} M.J. Martins, Phys. Rev. Lett., 74 (1995) 3316; Nucl. Phys. B, 450 (1995) 768; Phys. Lett. B, 359 (1995) 334.
\bibitem{BATCHELOR} M.T. Batchelor, J. de Gier, J. Links, M. Maslen, J. Phys. A: Math. Gen., 33 (2000) L97.
\bibitem{MARTINS-SP2N} M.J. Martins, J. Phys. A: Math. Gen. 35 (2002) L261; M.J. Martins, Nucl. Phys. B 636 (2002) 583.
\bibitem{GROUP} P. Ramond, Group theory: A physicist’s survey, Cambridge University Press, Cambridge, UK, 2010.
\bibitem{ODBA} Y. Wang, W. -L. Yang, J. Cao and K. Shi, Off-Diagonal Bethe Ansatz for Exactly
Solvable Models, Springer Press, 2015.
\bibitem{CDD} L. Castillejo, R. Dalitz, F. Dyson, Phys. Rev., 101 (1956) 543.
\bibitem{KLUMPER1990} A.~Kl\"umper, J. Phys. A: Math. Gen., 23 (1990) 809.
\bibitem{POLYNEG} V.S. Adamchik, J. Comp. Appl. Math., 100 (1998) 191.

\end{thebibliography}
\end{document}